\documentclass{article}
\usepackage{arxiv}

\usepackage[utf8]{inputenc} 
\usepackage[T1]{fontenc}    
\usepackage{natbib}
\usepackage{graphicx}
\usepackage{amsmath}
\usepackage{hyperref}

\usepackage{tabularx,booktabs,multirow,array}
\AtBeginDocument{%
  }

\usepackage{listings}
\usepackage{algorithm}
\usepackage{algorithmicx}
\usepackage{algpseudocode}

\newcommand{\benchmarkname}{CONCUR}

\begin{document}

\title{CONCUR: Benchmarking LLMs for Concurrent Code Generation}

\author{
    Jue Huang \\
    The University of Queensland \\
  \texttt{jue.huang@uq.edu.au} \\
    \And
 Tarek Mahmud \\
  Texas A\&M University - Kingsville\\
  \texttt{tarek.mahmud@tamuk.edu} \\
  \And
 Corina Pasareanu \\
  Carnegie Mellon University \\ 
  \texttt{pcorina@andrew.cmu.edu} \\
  \And
 Guowei Yang \\
  The University of Queensland \\
  \texttt{guowei.yang@uq.edu.au} \\
}



\maketitle

\begin{abstract}
Leveraging Large Language Models (LLMs) for code generation has increasingly emerged as a common practice in the domain of software engineering. Relevant benchmarks have been established to evaluate the code generation capabilities of LLMs. However, existing benchmarks focus primarily on sequential code, lacking the ability to effectively evaluate LLMs on concurrent code generation. Compared to sequential code, concurrent code exhibits greater complexity and possesses unique types of bugs, such as deadlocks and race conditions, that do not occur in sequential code. Therefore, a benchmark for evaluating sequential code generation cannot be useful for evaluating concurrent code generation with LLMs. To address this gap, we designed a benchmark \benchmarkname{} specifically aimed at evaluating the capability of LLMs to generate concurrent code. \benchmarkname{} consists of a base set of 43 concurrency problems derived from a standard concurrency textbook, together with 72 validated mutant variants, resulting in 115 total problems. The base problems serve as the semantic core of the benchmark, while the mutants expand linguistic and structural diversity. We conducted an evaluation of a range of LLMs on \benchmarkname{}, highlighting limitations of current models.

Overall, our work provides a novel direction for evaluating the capability of LLMs to generate code with focus on concurrency. 
\end{abstract}


\section{Introduction}
LLMs have demonstrated significant success across software engineering tasks, with code generation being one of their most common applications. This growing influence has spurred the creation of benchmarks to systematically evaluate model performance on program synthesis.
Although existing benchmarks \citep{chen2021evaluating,athiwaratkun2022multi,yu2024codereval} evaluate LLMs’ code generation capabilities and propose corresponding metrics and datasets, these benchmarks are primarily limited to sequential code.

Concurrency plays a crucial role in the software industry~\citep{hwu2008concurrency}. The development of concurrent programs is known to present numerous challenges~\citep{bianchi2017survey}. Unlike sequential code, concurrent programming is more complex due to the interleaved execution of tasks or threads~\citep{yu2009case}.  Concurrent code introduces a host of unique challenges, including nondeterministic thread scheduling, synchronization requirements, and subtle correctness issues such as race conditions, deadlocks, and starvation. These concurrency-specific bugs are notoriously difficult to anticipate, diagnose, and resolve, even for experienced developers. As a result, evaluating LLMs only on sequential code leaves a critical blind spot in understanding their true capabilities and limitations. This prevents many code generation benchmarks from reliably assessing the capability of LLMs to generate concurrent programs. Static similarity metrics, such as CodeBLEU~\citep{ren2020codebleu}, capture surface-level overlap and often overlook semantic correctness, while unit-test-based dynamic evaluation cannot systematically explore nondeterministic thread schedules. Therefore, these approaches may incorrectly classify flawed concurrent programs as correct.

To address this gap, we propose \benchmarkname{}, the first benchmark dedicated to evaluating LLMs on concurrent code generation. \benchmarkname{} is built around a base set of 43 carefully curated concurrency problems derived from a standard concurrency textbook~\citep{goetz2006java}. These base problems form the semantic core of the benchmark. To improve coverage and reduce the risk of memorization, we further construct 72 validated mutant variants of a subset of the base problems, resulting in 115 total problem instances used for evaluation. Unless otherwise stated, all experimental results are reported on the full set of 115 problems; we additionally report base-only results as a sanity check to ensure that our conclusions do not depend on the mutant variants. Beyond dataset design, \benchmarkname{} leverages formal methods techniques, namely model checking, to rigorously evaluate the correctness of the generated code.

We evaluate 23 state-of-the-art LLMs on \benchmarkname{}, including both proprietary APIs and open-source models. Our framework effectively detects concurrency bugs in LLM-generated programs, achieving 92\% precision in manual validation. Furthermore, our analysis reveals that the widely used code evaluation metric, CodeBLEU, fails to reliably reflect correctness of concurrent programs.

The key contributions of our paper are as follows.
\begin{itemize}
    \item We present a benchmark \benchmarkname{} for evaluating the capability of LLMs in concurrent code generation. It consists of 115 concurrency problems paired with structured prompts and validated ground-truth implementations in Java.

    \item \benchmarkname{} also introduces an automated validation framework using formal methods techniques, namely model checking.

    Unlike existing benchmarks that 
    rely on test cases to assess the correctness of code, 

    \benchmarkname{} uses model checking to exhaustively explore  the entire state space (within a user-defined bound) of the target concurrent program and verify its correctness. 

    \item We evaluate the capability of 23 LLMs in generating concurrent programs and analyze the potential issues inherent in the concurrent programs produced by these LLMs. The results indicate that our benchmark can effectively identify  weaknesses in current LLMs when producing concurrent code.
    \item We have made our dataset and associated tools publicly available at \url{https://anonymous.4open.science/r/CONCUR-9DD4}. We also maintain a public leaderboard at \url{https://concur-bench.github.io/concurbench.github.io/leaderboard.html}, where we will continue to add new LLMs and expand the benchmark with {additional concurrency problems}. 

\end{itemize}

\section{Benchmark}

To systematically evaluate LLMs on concurrent code generation, we introduce \benchmarkname{}, a benchmark that integrates a curated dataset of concurrency problems with a rigorous evaluation framework. The dataset consists of 115 concurrency problems, each paired with structured prompts and validated ground-truth implementations. Building on this foundation, the evaluation framework combines compilation through Java 8 compiler with model checking using Java Pathfinder (JPF)~\citep{havelund2000model}.

\subsection{Dataset} \label{sec:dataset}

To evaluate the ability of LLMs to generate concurrent programs, we constructed a dataset of concurrency problems paired with carefully designed prompts and ground-truth implementations. The dataset is designed to ensure that generated outputs can be evaluated consistently and reliably through compilation and model checking. 

\begin{figure}[b!]
\centering
\fbox{%
\begin{minipage}{0.85\textwidth}
Please write a concurrent program in \textbf{[Programming language]} that implements the following functionality:

\textbf{[Concurrency problem description]}.

The requirements are as follows:

\textbf{[Specifications]}
\end{minipage}
}
\caption{Prompt for concurrent program generation by LLMs.}
\label{fig: Structure for prompt}
\end{figure}

\subsubsection{Problem selection}
Two authors systematically reviewed the reference book~\citep{goetz2006java} and selected concurrency problems to form the basis of the benchmark. Problems that required third-party library packages or could not be validated through back-end output, such as tasks involving database operations or user interface interactions, were excluded. For problems involving a large number of objects, the descriptions were adjusted to reduce object counts. This ensured that LLMs would not produce solutions spawning an excessive number of threads, which would otherwise inflate the state space during model checking and increase execution time. The adjusted problems were then incorporated into the dataset.

Each selected problem was reformulated into a self-contained description suitable for prompting LLMs. The descriptions emphasized concurrency-specific requirements, such as synchronization behavior and inter-thread interactions, while bounding the number of threads and iterations to keep state space exploration tractable. Special care was taken to ensure that these descriptions preserved the spirit of the original problem while also being executable in our evaluation framework. In particular, we explicitly stated input-output behavior and concurrency requirements that might otherwise be implicit in the reference material.

\subsubsection{Prompt engineering}
To ensure consistency and reduce noise in generation, we designed a structured prompt (Figure~\ref{fig: Structure for prompt}) with three components: the programming language, the concurrency problem description, and the specification. In our benchmark, Java 8 was selected as the target language due to its maturity and strong support in analysis tools such as JPF. The framework enforces several constraints: (1) implementations must comply with Java 8 and avoid third-party libraries; (2) thread counts and iterations must be bounded to prevent state explosion during JPF analysis; (3) all code must be consolidated into a single block for compilation; and (4) a public class with a \texttt{main} method must be provided as the entry point.

These rules address three practical considerations. First, \emph{concurrent code characteristics:} many problem statements omit explicit object or thread counts, leading LLMs to generate code with excessive threads. Bounding these values mitigates state explosion and improves efficiency. Second, \emph{programming language constraints:} Java 8 prohibits duplicate class names and does not support third-party libraries in our setup. Prompts therefore forbid external imports and enforce unique class definitions to prevent compilation errors. Third, \emph{LLM output behaviors:} LLMs often interleave code with explanations or split solutions across multiple blocks, complicating extraction. To avoid this, prompts explicitly instruct models to output “all classes and methods within a single .java file,” ensuring a coherent and compilable program.

\subsubsection{Ground-truth construction}
The reference book provided only incomplete code snippets rather than runnable solutions. To address this, two authors reconstructed complete ground-truth programs by consolidating fragmented code into fully executable implementations. They prepared task lists for each problem, merged them through discussion, and resolved disagreements through consultation with a third author. Each ground-truth solution was extended with a \texttt{main} function to guarantee execution of all defined threads and to exercise inter-thread interactions. Thread counts and iteration limits were systematically adjusted to satisfy problem requirements while maintaining computational efficiency. The maximum path execution depth for each program was also determined and recorded as a reference for JPF configuration during evaluation.

\subsubsection{Mutant generation}
To increase the size and diversity of the benchmark, we additionally constructed 72 mutated variants derived from the 43 base problems. These mutations were generated by prompting the Gemini model to produce variations of the original problem descriptions. Each generated mutant was then manually validated to ensure that (1) it preserved the semantics of the original concurrency pattern, (2) it remained compatible with our evaluation framework, and (3) it introduced meaningful structural diversity. We generated three mutants per original problem; after manual validation, we retained 72 semantically correct mutants covering 24 of the 43 base problems.
These variations expand the benchmark to 115 total problems and also the mutated descriptions do not appear in the original textbook, reducing the likelihood of evaluation-time memorization.

The final dataset comprises 43 original concurrency problems and 72 validated mutants, resulting in 115 total task instances. The benchmark captures a representative spectrum of concurrency constructs and issues, including \texttt{synchronized} blocks, \texttt{volatile} variables, the \texttt{Lock} interface, \texttt{ReentrantLock}, \texttt{lock()/unlock()}, \texttt{tryLock()}, \texttt{Semaphore}, \texttt{CountDownLatch}, atomic classes (\texttt{Atomic*}), \texttt{BlockingQueue}, low-level threading with \texttt{Thread} and \texttt{Runnable}/\texttt{Callable}, and high-level task management via \texttt{ExecutorService}.

\subsubsection{Bounding Strategy for Model Checking}
\label{sec:bounding-strategy}
Systematic exploration of thread interleavings via model checking is inherently susceptible to state-space explosion. In concurrent programs, the number of reachable states grows rapidly with the number of threads, shared variables, and loop iterations. Without careful bounding, even small programs can become intractable for exhaustive exploration. To balance verification completeness with practical feasibility, \benchmarkname{} adopts an explicit bounding strategy that constrains concurrency scale while preserving the essential characteristics of each concurrency problem.

Our bounding strategy operates at two complementary levels: task design and verification configuration. At the task-design level, we explicitly bound the number of threads and the number of iterations for each concurrency problem. These bounds are incorporated into both the prompt specification and the corresponding ground-truth implementation. By aligning prompt constraints with the reference solution, we ensure that LLM-generated programs are evaluated under the same concurrency scale as the intended solution, rather than being penalized for exploring unnecessarily large execution spaces. Importantly, these bounds do not alter the underlying concurrency pattern (e.g., mutual exclusion, producer–consumer interaction, or concurrent read–write access), but merely limit the size of the explored state space.

At the verification level, we calibrate Java PathFinder (JPF) parameters using the ground-truth implementations. For each problem, we first execute JPF on the reference solution to determine its maximum execution depth under bounded exploration. We then set the \texttt{search.depth\_limit} for generated programs to ten times this reference depth, allowing substantially richer interleavings than those exercised by the ground truth while preventing unbounded growth. In addition, a fixed execution timeout is enforced through the \texttt{TimeLimitListener}, ensuring that pathological cases do not monopolize verification resources. This dual bounding—depth and time—provides a uniform and conservative safety net across all evaluated programs.

To illustrate the effect of bounding, consider a simple thread-safe mutable integer holder, where multiple threads increment a shared counter. In our dataset, both the ground truth and the prompt explicitly specify a small, fixed number of worker threads and a bounded number of increments per thread. Even under these modest bounds, JPF explores all relevant interleavings that could expose race conditions or synchronization errors. Increasing the number of threads or iterations would not introduce qualitatively new concurrency behaviors, but would dramatically increase the number of states to be explored. Thus, bounding enables tractable yet semantically meaningful verification.

This strategy reflects a deliberate trade-off. While unbounded verification would be ideal, it is infeasible in practice for concurrent programs of even moderate complexity. By grounding bounds in the behavior of reference implementations and applying them consistently across prompts and evaluations, \benchmarkname{} maintains fairness, scalability, and reproducibility without sacrificing the ability to detect concurrency-specific errors. Similar bounded exploration assumptions are common in practical model checking and testing of concurrent systems, and we follow this paradigm to enable large-scale benchmarking of LLM-generated concurrent code.

\subsection{Code Evaluation Framework} \label{benchmark-evaluation}
\begin{figure}[t!]
\begin{center}
\includegraphics[width=0.9\linewidth]{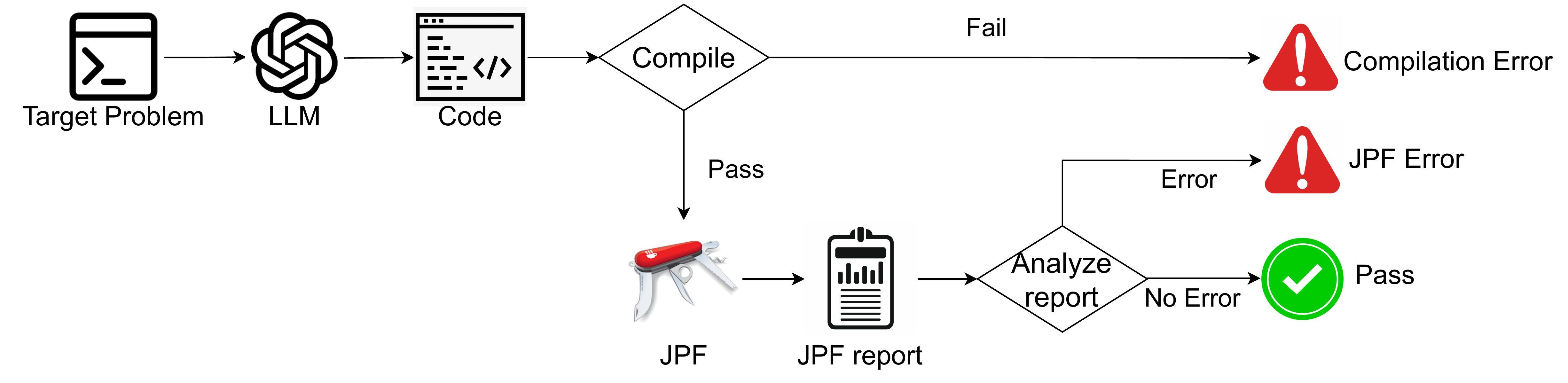}
\end{center}
\caption{Code Evaluation Framework of \benchmarkname{}.}
\label{fig: Whole process for code evaluation}
\end{figure}

Figure~\ref{fig: Whole process for code evaluation} illustrates the workflow of our evaluation framework. Starting from carefully designed prompts, LLMs generate candidate solutions from which we automatically extract Java code. To assess correctness, our framework combines both compilation in a controlled JDK environment and model checking with JPF. This dual-stage strategy allows us to detect errors that either prevent compilation or emerge only during concurrent execution.

Extracting clean, runnable code from raw LLM outputs is non-trivial, since responses may include interleaved text, explanations, or fragmented snippets. To automate this process, we leverage Java’s syntactic structure to identify keywords, encapsulate code into a single coherent file, and prepare it for evaluation. We also account for Java 8 syntax limitations to minimize false negatives caused by environment-specific constraints.

\subsubsection{Compilation in the JDK environment} 
As specified in the prompt framework, all problems must be solved using only default Java 8 libraries, without reliance on third-party packages. After generation, each Java file is named according to the declared \texttt{public class} and designated as the entry point for JPF analysis. This step ensures consistency across models and prevents misalignment between file names and class names.

Compilation errors are categorized into three main types:
\begin{itemize}
    \item {Missing package} – the program requires an import statement that is absent from the code.
    \item {Syntax error} – the program violates Java 8 syntactic rules.
    \item {Third-party dependency} – the program attempts to import unsupported external libraries.
\end{itemize}
To avoid duplicate class name conflicts, only one generated program is compiled and tested at a time. Programs that fail at this stage are labeled as \textit{compilation-failed}. Programs that compile successfully are automatically assigned a JPF configuration file and advanced to the concurrency testing stage, where model checking is performed. 

\begin{table}[t]
\caption{Error Types Detected by JPF}
\label{jpf-error-table}
\centering
\small
\begin{tabularx}{\columnwidth}{lX}
\toprule
\textbf{Error Type} & \textbf{Description} \\
\midrule
Deadlock (DL) &
A deadlock occurs when multiple threads are running and each thread waits indefinitely for others to release their locks. \\

Race Condition (RC) &
A race condition occurs when concurrent read and write operations access the same shared variable without proper synchronization. \\

Starvation (SV) &
Starvation occurs when one or more threads are perpetually denied access to shared resources because higher-priority threads monopolize execution. \\

Uncaught Exception (UE) &
Exceptions that arise during concurrent program execution, including both concurrency-related errors and all exceptions defined in the Java~8 standard libraries. \\

No Entry Method (NEM) &
The generated code lacks a public class with a \texttt{main} method, resulting in no valid entry point for execution. \\

Single Thread (ST) &
The generated code executes with only a single thread, which violates the prompt requirement for multi-threaded concurrent code. \\

Termination Error (TE) &
Errors occurring during JPF execution that originate from JPF itself or the runtime environment (e.g., resource exhaustion), causing premature termination. \\
\bottomrule
\end{tabularx}
\end{table}

\subsubsection{Model Checking with JPF} 

After filtering out programs with compilation errors, further analysis is required to detect potential concurrency bugs. For rigorous validation, we employ explicit-state model checking tool JPF. JPF systematically explores all possible thread interleavings, traversing execution paths to detect concurrency errors such as deadlocks, race conditions, and other defects listed in Table~\ref{jpf-error-table}.

JPF uses an extensible listener mechanism that enables fine-grained monitoring of program execution. We leverage default listeners to capture general runtime exceptions and extend them with custom listeners to track properties specific to our study. For example, although a single-threaded program may compile and execute without errors, it fails to satisfy our prompt requirement of generating concurrent code. To address this, we added a listener that reports the number of threads created during execution, allowing us to automatically detect single-threaded outputs. Similarly, we configure listeners to capture starvation and termination errors beyond those caught by default JPF checks.

A key limitation of JPF is the risk of state-space explosion when a program spawns a large number of threads or contains deep branching execution paths. To mitigate this, we bounded the number of threads and iterations in problem descriptions (Section~\ref{sec:dataset}) and set strict limits in the JPF configuration: the maximum execution depth is capped at ten times that of the corresponding ground truth. Ground-truth solutions were used to calibrate these parameters, ensuring that analysis remains tractable while preserving thorough exploration.

After execution, JPF produces detailed reports, which we parse to detect any errors listed in Table~\ref{jpf-error-table}. A generated program is considered correct only if it compiles successfully and no errors are reported by JPF.

\subsubsection{JPF Configuration and Listener Setup}
\label{sec:jpf-config}
To enable systematic and reproducible verification of LLM-generated concurrent programs, we configure Java PathFinder (JPF) with a unified execution environment and a fixed set of listeners tailored to concurrency analysis. For each Java program that successfully compiles, our framework automatically generates a corresponding JPF configuration file, ensuring that all models are evaluated under identical verification conditions.

Figure~\ref{fig:jpf-configuration} illustrates an example configuration. The \texttt{target} field specifies the public class containing the \texttt{main} method as the entry point for model checking. To control verification cost while maintaining adequate coverage of thread interleavings, we enforce both time and depth bounds. Specifically, we enable the \texttt{TimeLimitListener} to terminate exploration once a predefined timeout is reached, and we set \texttt{search.depth\_limit} to ten times the maximum execution depth observed in the corresponding ground-truth implementation. This heuristic allows JPF to explore substantially more interleavings than those exercised by the reference program, while preventing unbounded state-space explosion.

To detect a broad range of concurrency-related errors, we activate the following JPF listeners for every generated program:

\begin{itemize}
\item \textbf{ThreadCountListener} records the number of threads created during execution. This listener enables us to identify programs that compile and execute correctly but fail to spawn multiple threads, which violates the concurrency requirement of our benchmark.
\item \textbf{TimeLimitListener} enforces a uniform execution timeout, ensuring fairness across models and preventing pathological programs from monopolizing verification resources.
\item \textbf{StarvationListener} detects starvation scenarios by reporting threads that are never scheduled from creation to termination.
\item \textbf{PreciseRaceDetector} identifies potential data races by monitoring unsynchronized conflicting accesses to shared variables. This listener provides explicit race detection beyond JPF’s default exception-based checks.
\item \textbf{DeadlockAnalyzer} detects deadlocks by identifying cyclic lock dependencies among threads during execution.
\end{itemize}

Together, these listeners allow our framework to capture both classical concurrency bugs (e.g., deadlocks and race conditions) and benchmark-specific violations (e.g., single-threaded execution). All detected issues are reported in JPF’s verification output and subsequently parsed into the error categories summarized in Table~\ref{jpf-error-table}. Programs that terminate without triggering any listener-reported errors are considered to have passed the JPF verification stage and are labeled as correct by the automated oracle.

\begin{figure}[ht]
\centering
\fbox{%
\begin{minipage}{0.65\textwidth}
\ttfamily\raggedright
target=Main\\[0.3em]
+vm.fast.startup\\
listener=gov.nasa.jpf.listener.ThreadCountListener,\\
gov.nasa.jpf.listener.TimeLimitListener,\\
gov.nasa.jpf.listener.StarvationListener,\\
gov.nasa.jpf.listener.PreciseRaceDetector,\\
gov.nasa.jpf.listener.DeadlockAnalyzer\\[0.3em]
timeLimitMillis=9000\\[0.3em]
search.depth\_limit = 360
\end{minipage}%
}
\caption{Example JPF configuration.}
\label{fig:jpf-configuration}
\end{figure}

\subsubsection{From JPF Reports to Error Labels}

\begin{figure}[t]
\centering
\footnotesize
\setlength{\fboxsep}{4pt}
\fbox{%
\begin{minipage}{0.55\linewidth}

\textbf{System under test:} \texttt{VolatileCache.main()}\\
\textbf{Detected error:} \texttt{NotDeadlockedProperty}\\
\textbf{Summary:} Deadlock encountered during execution\\[0.6em]
\textbf{Blocked threads:}
\begin{itemize}
  \item \texttt{pool-1-thread-1} (WAITING on \texttt{LockSupport.park})
  \item \texttt{pool-1-thread-2} (WAITING on \texttt{LockSupport.park})
\end{itemize}
\textbf{Call stack (excerpt):}\\
\texttt{LockSupport.park()}\\
\texttt{AbstractQueuedSynchronizer\$ConditionObject.await()}\\
\texttt{LinkedBlockingQueue.take()}\\
\texttt{ThreadPoolExecutor.getTask()}\\
\texttt{ThreadPoolExecutor.runWorker()}\\[0.6em]
\textbf{Execution statistics:}
\begin{itemize}
  \item Max depth explored: 26
  \item States explored: 26
  \item Instructions executed: 19{,}441
  \item Unique threads created: 3
\end{itemize}
\textbf{Result:} Execution terminated with a deadlock.
\end{minipage}}
\caption{Example JPF report showing a detected deadlock in an LLM-generated concurrent program.}
\label{fig:example_jpf_report}
\end{figure}

After JPF execution completes, the verification results are summarized in a structured report that records detected property violations, execution traces, and exploration statistics. Figure~\ref{fig:example_jpf_report} shows a representative excerpt of a JPF report generated by our framework. The \texttt{results} section enumerates all violated properties reported by enabled listeners, while additional sections provide diagnostic information such as the triggering execution path, stack traces, and scheduling decisions.

Our evaluation framework automatically parses JPF reports and maps reported violations to a fixed set of error categories, as defined in Table~\ref{jpf-error-table}. This mapping is deterministic and rule-based, ensuring consistency across models and runs. Each listener produces distinct markers in the report, which allows us to reliably identify the corresponding error type. For example, deadlocks are detected through the \texttt{DeadlockAnalyzer} listener, which reports cyclic lock dependencies; data races are flagged by the \texttt{PreciseRaceDetector} listener when unsynchronized conflicting accesses to shared variables are observed; and starvation events are identified by the \texttt{StarvationListener} when a thread is never scheduled during execution.

In addition to concurrency-specific errors, our parser also captures uncaught runtime exceptions reported by JPF. These include both exceptions commonly associated with concurrent execution (e.g., \texttt{ConcurrentModificationException} and \texttt{RejectedExecutionException}) and general runtime exceptions that arise during multi-threaded execution (e.g., \texttt{NullPointerException} and \texttt{IllegalArgumentException}). All such cases are uniformly labeled as \emph{Uncaught Exception (UE)}, reflecting that the generated program terminates abnormally under at least one explored thread interleaving.

Certain violations correspond to benchmark-specific specification requirements rather than classical concurrency bugs. In particular, programs that compile and execute without raising concurrency-related exceptions may still fail to satisfy the requirement of generating concurrent code. To capture this class of errors, we label executions that create only a single thread as \emph{Single Thread (ST)} errors based on the output of the \texttt{ThreadCountListener}. Similarly, programs that lack a valid entry point (i.e., no public class with a \texttt{main} method) are labeled as \emph{No Entry Method (NEM)}. Finally, executions that terminate prematurely due to JPF-internal failures or resource exhaustion are categorized as \emph{Termination Errors (TE)}.

When multiple violations are reported for a single program, we record all detected error types to support fine-grained analysis of failure modes. However, for the purpose of pass/fail evaluation, a program is considered correct only if it compiles successfully and JPF reports no violations of any kind. This conservative criterion ensures that programs labeled as correct satisfy both syntactic requirements and concurrency-related specifications under all explored interleavings. 

\subsubsection{Detecting Non-Concurrent Solutions (ST)}

\begin{figure}[ht]
\centering
\scalebox{0.8}{
\fbox{%
\begin{minipage}{.92\textwidth}
\ttfamily\raggedright
import java.util.*;\\
import java.util.concurrent.*;\\
import java.lang.ref.*;\\[0.5em]

public class DelegatingVehicleTracker \{\\
\ \ private final ConcurrentHashMap<String, String> locations = new ConcurrentHashMap<>();\\[0.5em]

\ \ public DelegatingVehicleTracker(Map<String, String> initialLocations) \{\\
\ \ \ \ this.locations.putAll(initialLocations);\\
\ \ \}\\[0.5em]

\ \ public Optional<String> getLocation(String vehicleId) \{\\
\ \ \ \ return Optional.ofNullable(this.locations.get(vehicleId));\\
\ \ \}\\[0.5em]

\ \ public boolean setLocation(String vehicleId, String location) \{\\
\ \ \ \ if (!this.locations.containsKey(vehicleId)) \{\\
\ \ \ \ \ \ return false;\\
\ \ \ \ \} else \{\\
\ \ \ \ \ \ this.locations.put(vehicleId, location);\\
\ \ \ \ \ \ return true;\\
\ \ \ \ \}\\
\ \ \}\\[0.5em]

\ \ public Map<String, String> getAllLocations() \{\\
\ \ \ \ return Collections.unmodifiableMap(new HashMap<>(this.locations));\\
\ \ \}\\[0.5em]

\ \ public static void main(String[] args) throws Exception \{\\
\ \ \ \ Map<String, String> initialLocations = new HashMap<>();\\
\ \ \ \ initialLocations.put(``1", ``San Francisco");\\
\ \ \ \ initialLocations.put(``2", ``New York");\\
\ \ \ \ DelegatingVehicleTracker tracker = new \\
\ \ \ \ \ \ \ \ \ \ \ \ \ \ \ \ \ \ \ \ \ \ \ \ \ \ \ DelegatingVehicleTracker(initialLocations);\\
\ \ \}\\
\}
\end{minipage}}%
}
\caption{Thread-safe vehicle tracker implemented with \texttt{ConcurrentHashMap} that does not exhibit concurrent execution.}
\label{fig:single-thread-no-jpf-error}
\end{figure}

While Java PathFinder is effective at detecting classical concurrency bugs such as deadlocks and data races, it does not, by default, distinguish between truly concurrent programs and sequential executions that merely employ thread-safe abstractions. In the context of concurrent code generation, this distinction is critical: a program may compile, execute without errors, and even use concurrency-related data structures, yet still fail to exhibit any concurrent behavior.

In our benchmark, each task explicitly requires the generation of a concurrent program, meaning that multiple threads must be created and execute overlapping operations that potentially interact through shared state. However, we observed that LLMs frequently generate solutions that rely on thread-safe classes (e.g., \texttt{ConcurrentHashMap}, atomic variables, or synchronized methods) while executing all operations within a single thread. Such programs are often logically sound as sequential implementations, but they do not satisfy the concurrency requirements specified in the prompt.

To systematically identify this class of violations, we introduce the \emph{Single Thread (ST)} error category. Using the \texttt{ThreadCountListener} described in Section~\ref{sec:jpf-config}, we record the number of threads created during program execution under JPF. If execution involves only the main thread and no additional worker threads, the program is labeled as an ST error, even if no other concurrency-related violations are reported. This criterion ensures that programs labeled as correct not only avoid concurrency bugs, but also actively demonstrate concurrent execution.

Figure~\ref{fig:single-thread-no-jpf-error} illustrates a representative ST case. In this example, the generated program defines a thread-safe vehicle tracker using \texttt{ConcurrentHashMap}. Although the implementation is safe for concurrent access, the \texttt{main} method performs only initialization and does not spawn any additional threads. As a result, the program executes sequentially and produces no JPF-reported errors, yet it fails to meet the benchmark’s requirement of exercising concurrency. Without explicit detection of this pattern, such programs would be incorrectly classified as correct.

By explicitly distinguishing ST violations from other error types, our framework captures a common failure mode of LLM-generated code: conceptual awareness of concurrency constructs without corresponding concurrent execution. This distinction is particularly important for benchmarking, as it separates programs that are incorrect due to concurrency bugs from those that never realize concurrency at all. In our experiments, ST errors account for a non-negligible portion of failures (Table~\ref{tab:jpf_errors_pass1}), highlighting the necessity of treating non-concurrent solutions as first-class specification violations rather than overlooking them as benign cases.

\section{Experiments}
\subsection{Experimental setup}
The goal of our experiments was to systematically evaluate large language models (LLMs) on the task of concurrent code generation using the \benchmarkname{} benchmark. We began by benchmarking 23 LLMs, including both widely used APIs~\citep{claude-opus-4.1,gpt-5} and large open-source models~\citep{yang2025qwen3,dubey2024llama}. To ensure comparability, we primarily selected models with more than 30B parameters; in cases where only smaller variants were available, we used the largest publicly released version. Each model was prompted to solve all 115 concurrency problems in our dataset, following the structured framework described in Section \ref{benchmark-evaluation}. The outputs were compiled in a controlled Java 8 environment.

To better understand where models fail, we conducted a two-stage error analysis. In the first stage, we examined compilation errors, which were categorized into common causes such as missing imports, syntax violations, and the use of unsupported third-party libraries. In the second stage, we analyzed programs that compiled successfully using Java Pathfinder (JPF). 
This separation between compilation errors and JPF-detected errors enabled us to distinguish superficial problems of syntactic correctness from deeper issues related to concurrency.

Finally, we investigated how reflective CodeBLEU is as an evaluation metric in this setting. While CodeBLEU has been widely used as a static measure of code similarity, it is unclear whether high CodeBLEU scores correlate with dynamically correct concurrent programs. To address this, we compared CodeBLEU scores against ground-truth implementations and contrasted them with correctness outcomes established through JPF evaluation. This analysis allowed us to assess whether CodeBLEU provides meaningful insights in the context of concurrent code generation or whether dynamic evaluation remains indispensable.

\subsubsection{Implementation Details and Reproducibility Settings}

To ensure reproducibility and fair comparison across models, all experiments were conducted under a controlled and uniform execution environment. Generated programs were compiled using a fixed Java~8 development kit, and all verification runs were performed with the same version of Java PathFinder (JPF) and identical listener configurations, as described in Section~\ref{sec:jpf-config}. Each generated program was evaluated independently to avoid interference between executions, and no results were reused across different models or prompts.

For code generation, we use a consistent prompting strategy across all evaluated LLMs. Each model receives the same structured prompt for a given task, including the problem description, specification constraints, and explicit bounds on the number of threads and iterations. For pass@$k$ evaluation, $k$ responses are generated independently per prompt. No post hoc selection or ranking is performed beyond standard pass@$k$ semantics: a task is considered solved if at least one of the $k$ generated programs passes compilation and JPF verification. This setting reflects common usage scenarios of LLMs for code generation and avoids introducing additional heuristics that could bias results.

To reduce variance introduced by output formatting differences, we apply a uniform code extraction procedure to all model outputs. Specifically, we extract the Java program by identifying the public class declaration and retaining all code contained within the corresponding block, discarding any surrounding explanations or formatting artifacts. Outputs that do not contain a valid Java class definition are treated as compilation failures. This extraction strategy is applied consistently across models to minimize bias due to response style rather than code quality.

Compilation and model checking are executed with fixed resource limits. Each compilation is subject to a predefined timeout, after which the program is marked as compilation-failed. For JPF verification, we enforce both time and depth bounds using the \texttt{TimeLimitListener} and \texttt{search.depth\_limit} configuration. These limits are calibrated based on the corresponding ground-truth implementations, ensuring that generated programs are afforded sufficient exploration depth while preventing excessive resource consumption. Programs that exceed these bounds and terminate prematurely due to JPF-internal errors are labeled as termination errors.

All experiments are fully automated through a unified evaluation pipeline that orchestrates prompting, code extraction, compilation, JPF verification, and result aggregation. The pipeline records detailed logs for each stage, including compilation diagnostics, JPF reports, and execution statistics. Together with the released dataset, prompts, and configuration files, these logs enable independent reproduction of our results and facilitate further analysis of LLM behavior on concurrent code generation tasks.

\subsection{Results}

\begin{table}[t]
\caption{Comparison of 23 Large Language Models on Concurrent Code Generation for 115 Problems (Pass@1 vs.\ Pass@3).}
\label{tab:llm_comparison_pass1_pass3}
\centering
\small
\setlength{\tabcolsep}{8pt}

\newcolumntype{Y}{>{\raggedright\arraybackslash}X}

\begin{tabularx}{\columnwidth}{@{}Ycccc@{}}
\toprule
\multirow{2}{*}{\textbf{Model}} &
\multicolumn{2}{c}{\textbf{Passing Rate}} &
\multicolumn{2}{c}{\textbf{CodeBLEU}} \\
\cmidrule(lr){2-3} \cmidrule(lr){4-5}
& \textbf{k=1} & \textbf{k=3} & \textbf{k=1} & \textbf{k=3} \\
\midrule
gpt-5                    & 89/115 (77.39\%) & 105/115 (91.30\%) & 0.576122 & 0.576229 \\
claude-opus-4-1 & 78/115 (67.83\%) & 91/115 (79.13\%)  & 0.579037 & 0.580484 \\
gemini-3-pro             & 76/115 (66.09\%) & 85/115 (73.91\%)  & 0.574317 & 0.575100 \\
gpt-4o                   & 69/115 (60.00\%) & 95/115 (82.61\%)  & 0.547317 & 0.541937 \\
llama3.3:70b             & 57/115 (49.57\%) & 71/115 (61.74\%)  & 0.514043 & 0.514464 \\
qwen3:32b                & 53/115 (46.09\%) & 75/115 (65.22\%)  & 0.513917 & 0.509960 \\
wizardcoder:33b          & 48/115 (41.74\%) & 72/115 (62.61\%)  & 0.458396 & 0.460315 \\
codestral:22b            & 46/115 (40.00\%) & 67/115 (58.26\%)  & 0.469580 & 0.467794 \\
deepseek-r1:32b          & 43/115 (37.39\%) & 74/115 (64.35\%)  & 0.533301 & 0.536673 \\
phi4:14b                 & 39/115 (33.91\%) & 68/115 (59.13\%)  & 0.532162 & 0.531946 \\
phind-codellama:34b      & 38/115 (33.04\%) & 70/115 (60.87\%)  & 0.451454 & 0.446808 \\
opencoder:8b             & 36/115 (31.30\%) & 77/115 (66.96\%)  & 0.477348 & 0.476352 \\
gemma2:27b               & 34/115 (29.57\%) & 55/115 (47.83\%)  & 0.485571 & 0.485590 \\
mixtral:8x7b             & 33/115 (28.70\%) & 53/115 (46.09\%)  & 0.471824 & 0.476205 \\
codeqwen:7b              & 25/115 (21.74\%) & 54/115 (46.96\%)  & 0.469296 & 0.460804 \\
dolphin3:8b              & 22/115 (19.13\%) & 44/115 (38.26\%)  & 0.494357 & 0.490770 \\
magicoder:7b             & 12/115 (10.43\%) & 27/115 (23.48\%)  & 0.416781 & 0.412591 \\
llava:34b                & 7/115  (6.09\%)  & 20/115 (17.39\%)  & 0.451348 & 0.452500 \\
codellama:34b            & 7/115  (6.09\%)  & 23/115 (20.00\%)  & 0.402659 & 0.397490 \\
vicuna:33b               & 6/115  (5.22\%)  & 20/115 (17.39\%)  & 0.464678 & 0.452348 \\
starcoder2:15b           & 4/115  (3.48\%)  & 15/115 (13.04\%)  & 0.357170 & 0.356557 \\
zephyr:7b                & 4/115  (3.48\%)  & 14/115 (12.17\%)  & 0.530631 & 0.537083 \\
mistral:7b               & 3/115  (2.61\%)  & 14/115 (12.17\%)  & 0.502507 & 0.502946 \\
\bottomrule
\end{tabularx}
\end{table}

{\bfseries Benchmarking.} 
We evaluated the correctness of generated programs using pass@k~\citep{chen2021evaluating}, where codes are considered correct if they both compile successfully and pass JPF verification without errors. For each model, we report results under $k \in {1, 3}$, meaning that one or three generations per prompt were considered. A JPF configuration file was automatically generated for each Java program that passed compilation, ensuring consistent evaluation against the ground-truth solutions.

Table~\ref{tab:llm_comparison_pass1_pass3} reveals substantial differences between the pass@1 and pass@3 settings across the 23 evaluated LLMs. Overall, all LLMs achieve higher passing rates under pass@3, demonstrating that the choice of k has a significant impact on the evaluation of concurrent code generation. High-performing LLMs such as gpt-5 and gpt-4o exhibit large improvements (from 77.39\% to 91.30\% and from 60.00\% to 82.61\%, respectively). Moreover, 15 LLMs show pass-rate increases exceeding 15 percentage points, indicating a broad sensitivity to multi-sample generation. The effect is even more pronounced among weaker LLMs: codeqwen:7b nearly doubles its accuracy (21.74\% → 46.96\%), while codellama:34b improves from 6.09\% to 20.00\%. These findings highlight that allowing multiple generations disproportionately benefits mid- and low-performing LLMs, and that pass@1 alone may substantially underestimate the practical capabilities of many LLMs in concurrent code generation.

To validate the reliability of our automated evaluation, we conducted a manual inspection of LLM-generated code that passed compilation and JPF verification in the pass@1 setting. From the 829 programs identified as correct, we randomly sampled 115 for manual review, ensuring all models and problems are covered. For each selected program, two authors independently constructed a checklist based on the concurrency specification, then merged their assessments through discussion. We examined from the \texttt{main} function whether values were transmitted correctly across threads and compared the generated code against the ground truth in terms of classes, functions, data flow, and outputs.

Out of the 115 manually inspected programs, 106 were confirmed correct. Among the remaining cases, 2 exhibited improperly terminated threads, 2 failed to implement thread interaction, 1 contained incomplete class definitions, and 4 passed incorrect variables during execution. Despite these issues, the observed precision of 92\% demonstrates that our evaluation framework produces results that are both consistent and reliable, establishing it as a practical benchmark for concurrent code generation.

\begin{table}[t]
\caption{Summary of Errors in Generated Code (Pass@1)}
\label{tab:jpf_errors_pass1}
\centering
\small
\setlength{\tabcolsep}{3pt}
\renewcommand{\arraystretch}{1.15}

\newcolumntype{Y}{>{\raggedright\arraybackslash}X}
\newcolumntype{M}{>{\centering\arraybackslash}p{1.8cm}} 
\newcolumntype{N}{>{\centering\arraybackslash}p{0.72cm}} 

\begin{tabularx}{\columnwidth}{@{}Y M N N N N N N N@{}}
\toprule
\multirow{2}{*}{\textbf{Model}} &
\multirow{2}{*}{\shortstack[c]{\textbf{Compilation}\\\textbf{Errors}}} &
\multicolumn{7}{c}{\textbf{JPF Errors}} \\
\cmidrule(lr){3-9}
& & \textbf{DL} & \textbf{RC} & \textbf{SV} & \textbf{UE} & \textbf{NEM} & \textbf{ST} & \textbf{TE} \\
\midrule
claude-opus-4-1          & 1   & 5 & 7 & 0 & 9  & 0 & 0 & 15 \\
codellama:34b            & 88  & 2 & 2 & 0 & 8  & 3 & 2 & 9  \\
codeqwen:7b              & 64  & 1 & 3 & 0 & 1  & 6 & 3 & 4  \\
codestral:22b            & 42  & 6 & 3 & 0 & 5  & 2 & 1 & 10 \\
deepseek-r1:32b          & 43  & 1 & 3 & 0 & 4  & 0 & 1 & 17 \\
dolphin3:8b              & 74  & 3 & 2 & 0 & 3  & 2 & 1 & 5  \\
gemini-3-pro             & 0   & 6 & 5 & 0 & 14 & 0 & 0 & 3  \\
gemma2:27b               & 59  & 5 & 6 & 0 & 3  & 2 & 2 & 4  \\
gpt-4o                   & 11  & 1 & 7 & 0 & 9  & 0 & 1 & 1  \\
gpt-5                    & 2   & 2 & 6 & 0 & 7  & 0 & 1 & 4  \\
llama3.3:70b             & 41  & 0 & 9 & 0 & 4  & 2 & 3 & 1  \\
llava:34b                & 103 & 2 & 0 & 0 & 0  & 2 & 0 & 1  \\
magicoder:7b             & 80  & 3 & 0 & 0 & 0  & 3 & 4 & 8  \\
mistral:7b               & 103 & 1 & 1 & 0 & 3  & 2 & 1 & 0  \\
mixtral:8x7b             & 66  & 3 & 1 & 0 & 3  & 1 & 3 & 2  \\
opencoder:8b             & 43  & 1 & 2 & 0 & 4  & 0 & 1 & 5  \\
phi4:14b                 & 34  & 2 & 5 & 0 & 5  & 1 & 0 & 6  \\
phind-codellama:34b      & 49  & 0 & 1 & 0 & 6  & 3 & 3 & 13 \\
qwen3:32b                & 28  & 1 & 6 & 0 & 3  & 2 & 4 & 16 \\
starcoder2:15b           & 62  & 0 & 0 & 0 & 0  & 2 & 7 & 1  \\
vicuna:33b               & 73  & 0 & 0 & 0 & 1  & 1 & 1 & 1  \\
wizardcoder:33b          & 29  & 2 & 2 & 0 & 4  & 1 & 2 & 4  \\
zephyr:7b                & 123 & 0 & 0 & 0 & 1  & 0 & 0 & 0  \\
\midrule
\textbf{Total} & \textbf{1279} & \textbf{68} & \textbf{77} & \textbf{0} & \textbf{214} & \textbf{74} & \textbf{23} & \textbf{81} \\
\bottomrule
\end{tabularx}
\end{table}

\begin{figure}[t!]
    \centering
    \begin{minipage}[t]{0.45\textwidth}
        \centering
        \includegraphics[width=\linewidth]{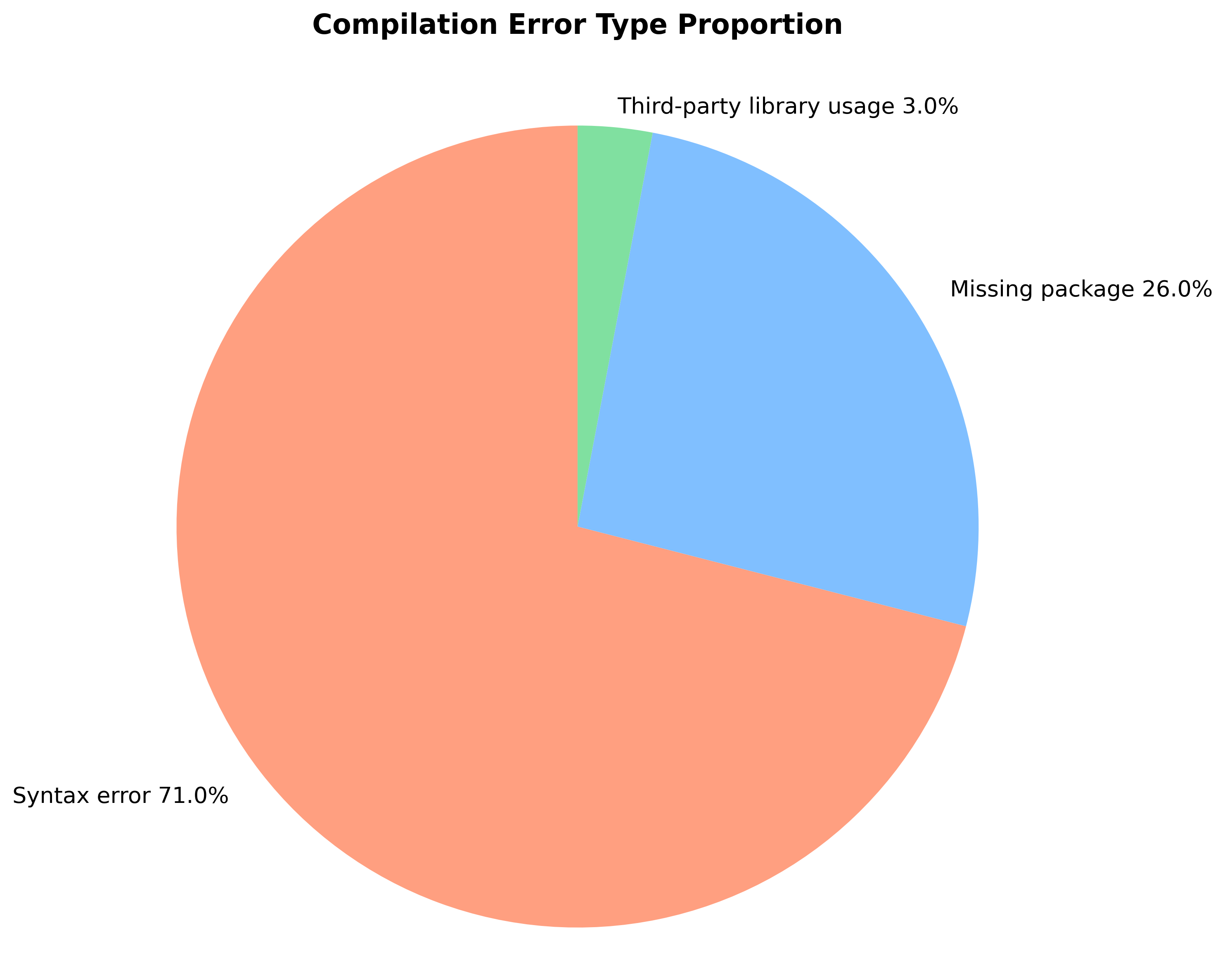}
        \caption{Distribution of compilation errors in 125 randomly selected generated codes.}
        \label{fig:Compilation failure distribution}
    \end{minipage}
    \hfill
    \begin{minipage}[t]{0.45\textwidth}
        \centering
        \includegraphics[width=\linewidth]{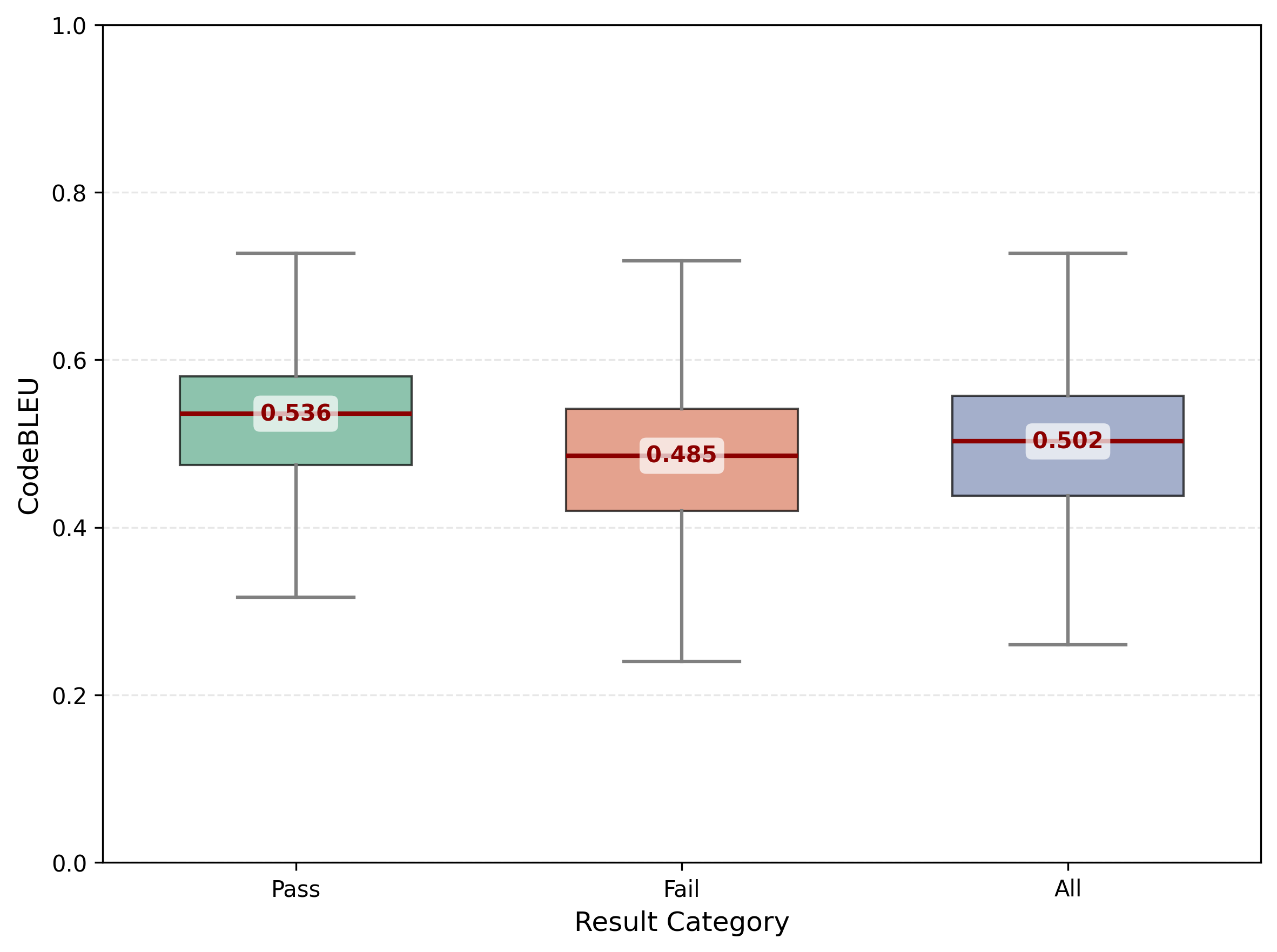}
        \caption{CodeBLEU score ranges for generated code.}
        \label{fig:boxplot}
    \end{minipage}
\end{figure}

{\bfseries Error analysis.} 
In this experiment, we focused on two categories of errors: compilation errors and JPF-detected errors.
At the compilation stage, the primary causes of errors were syntax errors and missing import statements. As illustrated in Figure~\ref{fig:Compilation failure distribution}, syntax errors accounted for 71.0\% of compilation failures, followed by missing package imports at 26.0\%. This finding aligns with observations in sequential code generation~\citep{fan2023automated}, where syntax errors remain a dominant source of errors. Our results underscore that syntax errors are significant across programming paradigms, including concurrent code generated by LLMs. 

Notably, 3.0\% of compilation failures stemmed from attempts to use third-party libraries, despite explicit instructions in the prompts to avoid such dependencies. This suggests that some LLMs occasionally prioritize generating seemingly convenient solutions over strictly adhering to task constraints, even when the original intent of the concurrency problems was to solve them using only fundamental language constructs and standard library classes. 

Moving beyond compilation, we analyzed JPF execution logs to categorize concurrency-related errors into the error types mentioned in Table~\ref{jpf-error-table}. Table~\ref{tab:jpf_errors_pass1} summarizes the JPF-reported errors under the pass@1 setting. Among these categories, uncaught exceptions emerged as the most frequent issue, indicating that LLMs often fail to anticipate and handle exceptions during execution. Despite prompts specifying that programs should include a public class and a main function, missing entry points still ranked among the most common errors, alongside termination issues arising from JDK version limitations and the evaluation environment.

Single-threaded execution was also a significant problem. Although these programs technically compiled and executed, they failed to satisfy the concurrency requirements of the tasks. We  manually inspected the 23 single-threaded programs identified, and 19 were found to be intended to design for concurrent access to support multi-threaded execution.

However, as the generated main thread did not instantiate multiple threads as required, these programs were ultimately classified as incorrect. This reflects a recurring pattern: LLMs often design concurrent logic conceptually but fail to implement the necessary threading constructs to realize actual parallel execution. Through an analysis of the statistics for the  concurrency issues, we found that race conditions and deadlocks occurred most frequently, indicating that LLMs are more prone to these types of errors when generating concurrent code. In contrast, starvation was not observed in any case. We hypothesize that this may be due to the relatively low complexity of the code, which allows each thread the opportunity to be scheduled.

A deeper review of JPF reports in the pass@1 setting revealed further limitations. Although JPF is capable of detecting exceptions defined in Java 8 libraries, certain concurrency-related exceptions, such as \textit{IllegalThreadStateException}, were absent from our reports. Two factors account for this: (1) the generated code did not cover the full spectrum of concurrency errors, and (2) LLMs appear reasonably effective at avoiding some of these exceptions when generating code. While our dataset and experiments did not exhaustively exercise every possible concurrency errors, \benchmarkname{} nonetheless demonstrates the ability to systematically detect uncaught exceptions supported by JPF.

In addition to the different error categories, we also collected secondary metrics from JPF, including maximum execution depth, total state space, and transition counts. However, we found that these attributes are highly sensitive to minor implementation details and do not correlate reliably with program correctness. Consequently, they are not used as evaluation metrics in our study. 

{\bfseries CodeBLEU analysis.}
In addition to JPF-based verification, we evaluated the similarity between generated code and ground‐truth implementations using CodeBLEU. The original CodeBLEU metric assesses code quality along four dimensions: n-gram overlap, weighted n-gram match, syntax, and data flow. Since our primary interest lies in logical correctness rather than surface-level resemblance, we focus on the syntax and data-flow components.

As shown in Table~\ref{tab:llm_comparison_pass1_pass3}, most LLMs obtain CodeBLEU scores between 0.45 and 0.58 for both k=1 and k=3. High-performing models such as gpt-5, claude-opus-4-1, and phi4:14b reach the upper end of this range, but several models exhibit clear mismatches between CodeBLEU and actual correctness. For example, mistral:7b and mixtral:8x7b achieve moderate CodeBLEU scores (0.47–0.50) despite very low pass rates. This indicates that higher syntax and data-flow similarity does not reliably correspond to correct concurrent behavior. Figure~\ref{fig:boxplot} further illustrates this limitation: correct programs have a higher median CodeBLEU (0.536) and a somewhat tighter distribution, yet there is substantial overlap with incorrect programs, some of which even score higher than correct ones. The mean CodeBLEU difference between all generated programs and only passing ones is also minimal, showing that CodeBLEU is only weakly correlated with correctness. 

Overall, while CodeBLEU captures certain surface-level similarities, it cannot serve as a standalone metric for evaluating concurrent code generation. Static similarity metrics are insufficient, underscoring the need for dynamic verification tools such as JPF. 

\subsubsection{Manual Audit Protocol for Passing Programs}

\begin{figure}[t!]
\begin{center}
\begin{verbatim}
        +---------------------------+        +---------------------------+
        |     Ground Truth          |        |     Generated Code        |
        +---------------------------+        +---------------------------+
        | Class: VehicleTracker     |        | Class: VehicleManager     |
        |---------------------------|        |---------------------------|
        | + getLocation()           |  <-->  | + fetchPosition()         |
        |   - return x, y           |        |   - return a, b           |
        |                           |        |                           |
        | + setLocation()           |  <-->  | + assignPosition()        |
        |   - set x, y              |        |   - set a, b              |
        |                           |        |                           |
        | + Location()              |  <-->  | + Position()              |
        |   - int x, y              |        |   - float x, y            |
        +---------------------------+        +---------------------------+
\end{verbatim}
\end{center}
\caption{Class and function comparison between ground truth and generated code.}
\label{fig:class_function_comparison}
\end{figure}

Although our automated evaluation framework based on compilation and JPF verification provides strong guarantees for detecting concurrency-related violations, it does not fully capture all aspects of functional correctness specified in the prompts. In particular, a program may avoid deadlocks, data races, and uncaught exceptions under all explored interleavings, yet still fail to implement the intended concurrent behavior or functional specification. To assess the reliability of our automated oracle and to characterize its limitations, we conducted a systematic manual audit of a subset of programs classified as correct.

We focused our manual inspection on programs that passed both compilation and JPF verification under the pass@1 setting. From this set, we randomly sampled 115 programs, ensuring coverage across all evaluated models and problem categories. Each sampled program was independently reviewed by two authors, following a structured audit protocol designed to minimize subjectivity and ensure consistency. Disagreements were resolved through discussion, with a third author consulted when consensus could not be reached.

The audit protocol proceeds in three stages. First, we verify structural alignment between the generated program and the ground-truth implementation. This step checks whether the required classes, methods, and shared data structures specified by the prompt are present in the generated code, allowing for benign renaming of identifiers and refactoring. Figure~\ref{fig:class_function_comparison} illustrates this alignment process, where semantically corresponding classes and methods are matched despite differences in naming or internal representation.

Second, we examine the \texttt{main} method to determine whether the program actually exercises concurrent execution as intended. This includes verifying that multiple threads are created, that they execute overlapping operations, and that they interact through shared objects rather than independent instances. Programs that instantiate separate objects per thread or fail to trigger concurrent interactions are flagged at this stage, even if they are free of concurrency-related runtime errors.

Third, we assess functional and concurrency-specific requirements using a prompt-derived checklist. For each task, we construct a to-do list capturing essential requirements such as the number of threads, the expected pattern of interaction (e.g., producer–consumer, concurrent read/write), and required synchronization mechanisms. A program is considered incorrect if any checklist item is violated, even when execution completes without errors. This step ensures that correctness judgments are grounded in the original specification rather than solely in the absence of detected bugs.

Using this protocol, 106 out of the 115 audited programs were confirmed to be correct, yielding a precision of 92\% for our automated evaluation pipeline. The remaining cases reveal systematic limitations of automated checking: two programs terminated threads prematurely, two failed to implement required thread interactions, one contained incomplete class definitions, and four propagated incorrect values across threads. These findings indicate that while JPF-based verification is effective for detecting concurrency errors, a small fraction of functional or specification-level violations may escape automated detection.

Overall, this manual audit demonstrates that our automated oracle provides a high-precision approximation of correctness for concurrent code generation, while also highlighting the boundaries of what can be reliably verified without explicit functional specifications or assertions. The identified failure modes motivate future extensions of the benchmark with richer oracles, such as automatically injected assertions or specification-driven checking.

\subsubsection{Representative False Positives Missed by Automated Checks}
\label{sec:false-positives}

\begin{figure}[t]
\centering
\begin{lstlisting}[language=Java]
public static void main(String[] args) {
  Thread t1 = new Thread(() -> {
    ServerStatusBeforeSplit ssb = new ServerStatusBeforeSplit(); // per-thread instance
    ...
  });
  Thread t2 = new Thread(() -> {
    ServerStatusBeforeSplit ssb = new ServerStatusBeforeSplit(); // per-thread instance
    ...
  });

  ExecutorService ex = Executors.newFixedThreadPool(2);
  ex.submit(t1); ex.submit(t2);
  ex.shutdown();
}
\end{lstlisting}
\caption{A representative false positive where threads do not share state despite concurrent execution.}
\label{fig:server-status-before-split}
\end{figure}

Although our automated evaluation pipeline achieves high precision in identifying correct concurrent programs, the manual audit reveals a small but non-negligible set of false positives. These cases pass compilation and JPF verification without reporting any violations, yet fail to satisfy the functional or concurrency-specific requirements stated in the prompt. In this section, we analyze representative examples to illustrate the limitations of automated checking and to characterize common failure patterns.

A prevalent false-positive pattern arises from the absence of meaningful thread interaction. Figure~\ref{fig:server-status-before-split} shows an LLM-generated implementation of the \texttt{ServerStatusBeforeSplit} task. Although the program explicitly creates multiple threads and executes without raising concurrency-related exceptions, each thread instantiates its own independent object rather than sharing a common instance. As a result, the threads do not interact through shared state, and the execution effectively degenerates into multiple independent sequential computations. From JPF’s perspective, this program is free of deadlocks, data races, and uncaught exceptions, and therefore passes verification. However, it violates the prompt’s requirement that concurrent threads coordinate through shared resources.

Another class of false positives involves incomplete realization of thread lifecycles. In several audited programs, threads are created but terminate prematurely or are not properly synchronized with the main thread. For example, worker threads may be started but not joined, or execution may proceed without ensuring that concurrent operations have completed before program termination. While such programs may still execute without triggering JPF-reported errors under bounded exploration, they fail to guarantee the intended concurrent behavior or output consistency specified in the task.

We also observed cases where values are incorrectly propagated across threads despite the absence of detected concurrency bugs. In these programs, synchronization mechanisms prevent data races, but logical errors—such as updating the wrong shared variable or passing incorrect arguments to worker threads—lead to incorrect outcomes. These errors are functional in nature and fall outside the scope of what JPF can detect without explicit assertions or specification-level properties.

These representative false positives highlight an important boundary of automated verification for concurrent code generation. JPF-based checking is well suited for detecting concurrency safety violations, but it cannot, by itself, ensure that generated programs faithfully implement the intended functional semantics of a task. Consequently, our benchmark adopts a conservative definition of correctness for automated evaluation while complementing it with manual auditing to estimate precision and to expose systematic gaps. Addressing these gaps remains an open challenge and motivates future work on integrating specification-aware or assertion-based oracles into the evaluation framework.

\subsection{Limitations}

While Java PathFinder (JPF) is effective in detecting a range of concurrency issues such as race conditions and deadlocks, it cannot capture all possible error types. For example, livelocks and  higher-level semantic violations remain undetected. JPF can check for assert violations but only if assertions are present in the code.  Further, we run JPF with depth and time bounds, to make the experimentation manageable. As a result, some errors may be missed due to the bound. This restricts the completeness of our concurrency-focused evaluation. Nevertheless, a manual inspection of 115 randomly selected correct programs indicated a recall of 92.2\%, suggesting that our approach can be reliable.
Furthermore, termination errors (TE) may be inconclusive (e.g., the program may or may not contain an error which might not be discovered due to JPF crashing). However, in this study, we consider this as an error since running JPF on the ground truth solutions completes.

The core problems originate from a widely used textbook, raising the possibility that models were exposed to related material during training. However, the textbook does not provide complete code for most problems; instead, it contains partial snippets that required extensive consolidation, restructuring, and augmentation to form executable programs. This limits the likelihood that an LLM could rely purely on memorization to produce correct solutions.

\section{Related Work}
{\bfseries LLMs in Code Generation and Benchmarking.}
Large language models (LLMs) have become central to software engineering tasks such as code generation~\citep{li2022competition,wang2023review,nejjar2025llms}, automated program repair~\citep{fan2023automated,joshi2023repair,xia2022less,zhang2024pydex,parasaram2024fact,xia2023automated}, and code translation~\citep{roziere2021leveraging}. Trained on massive code corpora, modern LLMs can synthesize complex structures efficiently, and many models are now widely used as coding assistants in both research and industry~\citep{codestral22b,roziere2023code,qwen,huang2024opencoder,luo2023wizardcoder}.
Despite these advances, their reliability remains limited. LLMs perform well on small, isolated functions but often struggle with end-to-end program generation, where correctness depends on reasoning across multiple components and execution contexts~\citep{li2022competition}.

To evaluate the capabilities of LLMs in code generation, a variety of benchmarks have been proposed~\citep{du2024evaluating,chen2021evaluating,liu2023your,yu2024codereval,austin2021program,sharma2025assessing,athiwaratkun2022multi,lai2023ds}. These benchmarks typically measure performance using datasets of prompts and ground-truth solutions, allowing comparisons across different models. However, most existing benchmarks emphasize unit- or function-level tasks and therefore do not fully capture the challenges of complex program synthesis. Furthermore, many rely exclusively on static analysis, which is insufficient for uncovering deeper semantic or runtime errors~\cite{al2010comparing}.

A particularly important gap lies in the evaluation of concurrent programs. Current benchmarks overwhelmingly target sequential code~\citep{hendrycks2021measuring,li2022competition}, leaving concurrency underexplored. This is problematic because concurrent programming is pervasive in modern applications, yet testing concurrency introduces significantly greater complexity than testing sequential code~\citep{taylor1992structural,sen2007effective}. Compilers can only ensure syntactic correctness, not semantic correctness, and runtime execution alone cannot be relied upon due to nondeterminism in thread scheduling~\citep{carver1998use,fonseca2011finding}. Thus, existing evaluation strategies are insufficient for benchmarking the generation of reliable multithreaded programs. 

{\bfseries Evaluation for code by static and dynamic analysis}
Traditional benchmarks~\citep{du2024evaluating,chen2021evaluating,liu2023your,yu2024codereval,austin2021program,sharma2025assessing,athiwaratkun2022multi,lai2023ds} employ both static and dynamic analysis. Static analysis often leverages similarity-based metrics, such as BLEU~\citep{papineni2002bleu}, CodeBLEU~\citep{ren2020codebleu}, or CrystalBLEU~\citep{eghbali2022crystalbleu}, to compare generated code against references. Compilation is also a necessary prerequisite for both static and dynamic evaluation, ensuring the generated code conforms to language specifications and enabling black-box testing methods such as fuzzing. Dynamic analysis, in turn, executes the generated code under controlled settings to verify functional correctness~\citep{miller1990empirical}.

For concurrent programs, however, these methods are insufficient. Limited runtime execution cannot capture all possible thread interleavings, leaving concurrency-specific issues like race conditions and deadlocks undetected. To address this, we employ Java PathFinder (JPF), a model checker that systematically explores thread schedules. By running generated code under JPF configurations and comparing results to ground truth extracted in advance, our framework enables precise detection of concurrency errors. This provides a rigorous basis for benchmarking LLMs on multithreaded program generation.

\section{Conclusion}
We propose a new benchmark, \benchmarkname{}, the first benchmark specifically designed to evaluate large language models on concurrent program generation. Unlike existing benchmarks that primarily target sequential code, \benchmarkname{} addresses the unique challenges of concurrency by providing a curated dataset of 115 problems and an evaluation framework that integrates compilation with systematic model checking. Through experiments on 23 state-of-the-art LLMs, we show that while many models can produce compilable code, they frequently fail to generate programs that are truly correct under all thread interleavings. Our analysis further demonstrates that static similarity metrics, such as CodeBLEU, do not reliably capture concurrency correctness, underscoring the necessity of the evaluation framework.

In future work, we plan to explore and adopt more concurrent path execution analysis frameworks to extend the benchmark for evaluating code generation across other programming languages. 

\section{Data Availability}
The dataset and associated tools are publicly available at \url{https://anonymous.
4open.science/r/CONCUR-9DD4}. We also maintain a public leaderboard at \url{https://concur-
bench.github.io/concurbench.github.io/leaderboard.html}, where we will continue to add
new LLMs and expand the benchmark with additional concurrency problems.

\bibliographystyle{plain}
\bibliography{Bibliography}

\end{document}